\documentclass[twocolumn,superscriptaddress,showpacs,prl,amsmath,amstex,amssymb,citeautoscript,longbibliography]{revtex4-1}
\pdfoutput=1
\usepackage{natbib}
\usepackage[english]{babel}
\usepackage{letltxmacro}
\usepackage{latexsym}
\LetLtxMacro{\ORIGselectlanguage}{\selectlanguage}
\makeatletter
\DeclareRobustCommand{\selectlanguage}[1]{%
  \@ifundefined{alias@\string#1}
    {\ORIGselectlanguage{#1}}
    {\begingroup\edef\x{\endgroup
       \noexpand\ORIGselectlanguage{\@nameuse{alias@#1}}}\x}%
}
\newcommand{\definelanguagealias}[2]{%
  \@namedef{alias@#1}{#2}%
}
\makeatother
\definelanguagealias{en}{english}
\definelanguagealias{English}{english}
\usepackage{graphicx}
\usepackage{amsmath}
\usepackage{amsfonts}
\usepackage{amssymb}
\usepackage{bm}
\usepackage{color}
\usepackage[percent]{overpic}
\usepackage{soul} 
\usepackage{amssymb}
\usepackage{wasysym}
\usepackage{dsfont}
\usepackage{float}

\usepackage{hyperref}
\hypersetup{
    bookmarks=false,         
    unicode=false,          
    pdftoolbar=false,        
    pdfmenubar=true,        
    pdffitwindow=false,     
    pdfstartview={FitH},    
    pdftitle={},    
    pdfauthor={Authors},     
    pdfsubject={},   
    pdfcreator={},   
    pdfproducer={}, 
    pdfkeywords={many-body localization} {matrix elements} {disordered systems}, 
    pdfnewwindow=true,      
    colorlinks=true,       
    linkcolor=black,          
    citecolor=blue,        
    filecolor=magenta,      
    urlcolor=blue           
}

\newcommand{\be}{\begin{equation}}
\newcommand{\ee}{\end{equation}}
\newcommand{\bea}{\begin{eqnarray}}
\newcommand{\eea}{\end{eqnarray}}

\usepackage{graphicx}
\usepackage[colorinlistoftodos]{todonotes}
\usepackage{verbatim}
\usepackage[normalem]{ulem}

\begin{document}

\title{Critical Time Crystals in Dipolar Systems
}
\author{Wen Wei Ho}
\affiliation{Department of Theoretical Physics, University of Geneva, 1211 Geneva, Switzerland  }

\author{Soonwon Choi}
\affiliation{Department of Physics, Harvard University, Cambridge, Massachusetts 02138, USA}

\author{Mikhail D. Lukin}
\affiliation{Department of Physics, Harvard University, Cambridge, Massachusetts 02138, USA}

\author{Dmitry A. Abanin}
\affiliation{Department of Theoretical Physics, University of Geneva, 1211 Geneva, Switzerland  }

\date{\today}
\begin{abstract}

We analyze the  quantum dynamics of periodically driven, disordered systems in the presence of long-range interactions. Focusing on the stability of discrete time crystalline (DTC) order in such systems, we  use a perturbative procedure to evaluate its lifetime.
For 3D systems with dipolar interactions, we show that the corresponding decay is parametrically slow, implying that robust, long-lived DTC order can be obtained. We further predict a sharp crossover from the stable DTC regime into a regime where DTC order is lost, reminiscent of a phase transition. These results  are in good agreement with the recent experiments utilizing a dense, dipolar spin ensemble in diamond [Nature 543, 221-225 (2017)]. They demonstrate the existence of a novel, {\it critical} DTC regime that is stabilized not by many-body localization but rather by slow, critical dynamics.  Our analysis shows that the DTC response can be used as a sensitive probe of 
nonequilibrium quantum matter.

\end{abstract}

\maketitle

%

\emph{Introduction.}---Understanding and controlling non-equilibrium quantum matter is an exciting frontier of physical science. While periodic driving has long been used to control the properties of quantum systems,  it was only recently realized that  periodically driven (Floquet) systems can also host new states of matter that are not possible in equilibrium. 
In particular, this is possible in a class of driven disordered systems exhibiting many-body localization (MBL)~\cite{nandkishore2015MBLreview}, called Floquet-MBL systems, which avoid unbounded heating to infinite temperature~\cite{Ponte15,Lazarides15,Abanin20161}. The latter is generally expected to befall all ergodic isolated systems due to external driving~\cite{Lazarides14,Ponte14,Alessio14}.

One remarkable example of a novel, nonequilibrium phase of matter is the recently introduced discrete time crystal (DTC) \cite{Khemani16,Else16,Curt16,Yao_dtc:2016wp}, which  is characterized by a spontaneously broken discrete time-translation symmetry of the underlying drive.
In such systems, physical observables exhibit robust oscillations with a period that is an integer multiple of the underlying driving period $T$.  Indeed, key signatures of such robust DTC order have been observed  in two recent experiments~\cite{Zhang2017,Choi16DTC}. In particular, one of these realizations involved a disordered  ensemble of $\sim 10^6$ spins  associated with nitrogen-vacancy (NV) centers in diamond, which interact between themselves via dipolar couplings~\cite{Choi16DTC}.
The origin of apparent robustness of the observed DTC order in such a system~\cite{Choi16DTC}, however, has not been fully understood. Although this system is disordered due to the random positions of the NV centers in   3D, the  long-range dipolar interactions are believed to preclude 
localization 
\cite{Anderson:2011wp, Levitov:1990zz,Burin06,Yao:2014jj,Gutman16}.
Moreover, a prethermal regime of the DTC~\cite{ElsePrethermal17}, 
was also ruled out~\cite{Choi16DTC} since in the experiment the initial polarized state is effectively at an infinite temperature with respect to the effective Hamiltonian due to the randomly varying signs of the dipolar interactions. Since neither localization nor prethermalization are likely the mechanisms that stabilize the DTC order, this raises important questions about the origin of the observed robust DTC response.


This Letter   develops a theoretical treatment of DTC order in systems with long-range interactions.  
We utilize a perturbative approach to  analyze the interplay of long-range interactions, periodic driving, and positional disorder of spins. 
Focusing on dipolar systems in 3D, we show that although DTC order is only transient, it can persist for asymptotically long times 
with a strongly suppressed  thermalization rate. This behavior is intrinsically connected to the slow thermalization  
dynamics of disordered dipolar systems in 3D, which has been previously shown to be consistent with the so-called critical  regime~\cite{Anderson:2011wp,kucsko} without a periodic drive. 
 As a function of experimental parameters, we find that the relaxation time shows a sharp crossover between a regime where the DTC response is robust and a regime where it decays rapidly. This crossover is reminiscent of a phase transition, thereby allowing us to  obtain the effective phase diagram of the DTC which is in good agreement with experimental results. Thus, our work provides an explanation of the recent experimental observations~\cite{Choi16DTC}, and also demonstrates the possibility of the  DTC in systems with critical dynamics, a regime which we refer to as  `critical time crystals'. Furthermore, our perturbative approach 
can be used to study the nonequilibrium properties in other driven disordered systems with long-range interactions.

\begin{figure}[h]
\center
\includegraphics[width=0.95\columnwidth]{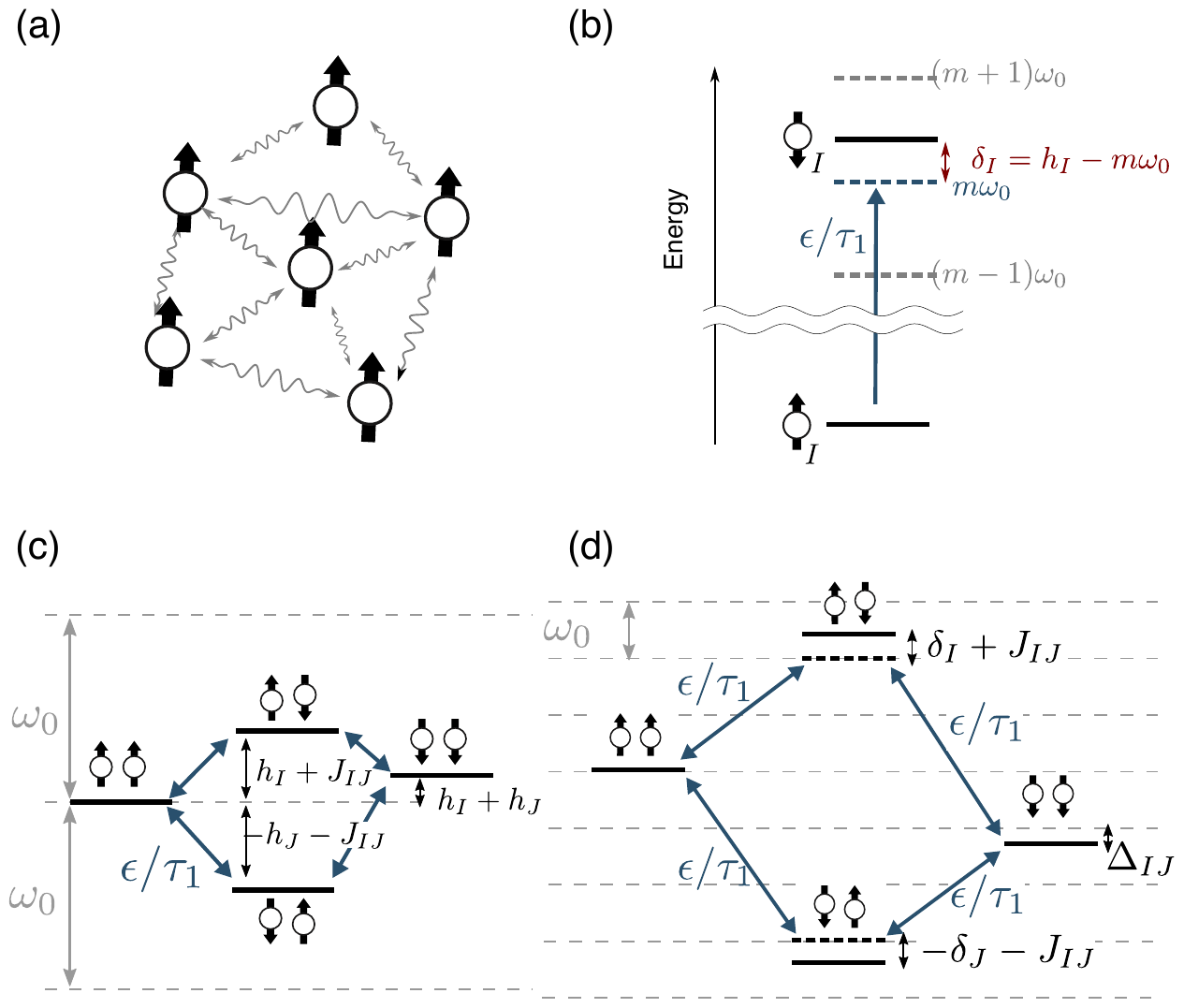}
\caption{(a) Ensemble of randomly positioned spins in 3D interacting via dipolar interactions. (b) Illustration of single-spin-flip processes. (c), (d) Energy level diagram for the second-order process of two spins flipping, in two regimes: (c) high frequencies ($\omega_0 \gg W$) and (d) low ($\omega_0 \ll W$) frequencies. The applied field flips a spin with magnitude $\epsilon/\tau_1$, which costs energy $\sim h_I - m^* \omega_0$.}
\label{fig:local}
\end{figure}

Our key results can be understood by considering a simple spin model that describes
an ensemble of dipolar interacting NV centers, used in the experiments of Ref.~\cite{Choi16DTC}.
Using strong microwave excitations, the effective Ising interactions between spins were  engineered, described by the following Hamiltonian:
\be\label{eq:H0}
H_{0}= \sum_i \Omega S^x_i+\sum_{i,j} \frac{J_{ij}}{r_{ij}^3} S_i^x S_j^x,
\ee 
where $\vec{S}_i = (S_i^x, S_i^y, S_i^z)$ are Pauli spin-1/2 operators, $\Omega$ the strong microwave driving along $\hat{x}$, $J_{ij}$ the orientation-dependent coefficient of dipolar interactions with typical strength $J_0$, and $r_{ij}$ the distance between spins $i$ and $j$.
We assume that the spin-1/2 particles are randomly distributed in three-dimensional space with density $n_0$ and neglect coupling to the environment  [Fig.~\ref{fig:local}(a)].
DTC order was observed by interrupting the evolution under Hamiltonian (\ref{eq:H0})  with rapid, global pulses that rotate the spin ensemble along the $\hat{y}$ axis by an angle $\pi + \epsilon$. The corresponding Floquet unitary is given by
\begin{align}
\label{eq:simplified_floquet}
U_F = \exp{\left[-i \sum_i (\pi + \epsilon) S_i^y\right]} \exp{\left[-i H_0 \tau_1 \right]},
\end{align}
where $\tau_1$ is the period for which the spins are allowed to interact for. In the experiment, the period is chosen such that $\Omega\tau_1=2\pi n$, and therefore $\Omega$ can effectively be taken to be $0$ in Eq.~(\ref{eq:H0}). 
When the system is initialized in a state where all spins are polarized along the $+\hat{x}$ direction, a nontrivial temporal response may be revealed by measuring the average polarization $P(n\tau_1)$ of the ensemble along $\hat{x}$ after $n$ Floquet cycles,
%
or equivalently, $q(n) \equiv (-1)^n P(n\tau_1)$, 
which serves as an order parameter for the DTC phase.  The stability of  the DTC order can be ascertained by studying the decay rate of $q(n)$ for large number of cycles as a function of $\tau_1$ and $\epsilon$.

In order to describe the dynamics of $q(n)$, we move into a so-called toggling frame, which rotates by $P_\pi \equiv \prod_j \exp{\left[ -i \pi S_j^y\right]} $ each time a global pulse is applied to the system.
Since $P_\pi S_i^x (P_\pi)^{-1} = -S_i^x$, the $2\tau_1$-periodic oscillation in $P(n\tau_1)$ naturally appears as a time-independent spin polarization in this new frame.
The dynamics of the system is then described by the Floquet unitary $\bar{U}_F = \exp{\left[-i \sum_i \epsilon S_i^y\right]} \exp{\left[-i H_0 \tau_1 \right]}$, or, equivalently, by an effective time-dependent Hamiltonian 
\begin{align}
H(t) & =  \sum_{ij}  \frac{J_{ij}}{r_{ij}^3} S_i^x S_j^x + \epsilon \sum_i S_i^y \sum_n \delta(t - n^{-} \tau_1).
\label{eqn:H}
\end{align}
Thus, our problem reduces to studying the depolarization dynamics of an initialized polarized spin ensemble under the time evolution of $H(t)$.
\emph{Physical picture.}---The essence of our analysis is to study resonant spin dynamics that lead to depolarization perturbatively in $\epsilon$, while accounting for energy exchanges provided by the external drive.
In particular, since $\sum_n \delta(t - n^{-} \tau_1) =  \frac{1}{\tau_1} \sum_m e^{i m\omega_0 t}$,
the pulsed periodic spin rotations can be viewed as spin excitation with harmonics of the fundamental frequency $\omega_0 \equiv 2\pi/\tau_1$ and fixed magnitude $\epsilon/\tau_1$.
While this driving allows energy absorption and emission in integer multiples of $\omega_0$, the interplay of strong interactions and positional disorder suppresses direct energy exchanges such that typical spins depolarize only via indirect higher-order processes in $\epsilon$.

%
%
Let us first consider the case without perturbations, i.e.~$\epsilon = 0$. Then the polarization of each spin along $\hat{x}$ is conserved. When all spins are initially polarized, each spin therefore experiences a mean-field potential $h_i \equiv \sum_{j \neq i} (J_{ij}/r_{ij}^3) \langle S_j^x \rangle$. 
Because of the random positioning of spins, the strength of $h_i$ is also random with zero mean and variance  $W^2 = \langle \frac{1}{4} (\sum_{j\neq i} J_{ij}/ r_{ij}^3)^2  \rangle$, where $\langle \cdot \rangle$ denotes averaging over different positions.

When $\epsilon \neq 0$, there is depolarization due to spins experiencing a time-varying on-site field along the $\hat{y}$ axis. Let us therefore consider the first-order process where spins individually flip due to the action of this field. If a spin experiences a a strong mean-field potential $h_i$ compared to the applied field, that is, if $h_i \tau_1 \gg \epsilon$, then it does not flip -- it experiences an 
effective field that is approximately pointing along the $\hat{x}$ axis and therefore precesses around it without significant depolarization.
On the other hand,  if $h_i \tau_1$ is close to an integer multiple of $2\pi$, then the spin rotates along  the $\hat{y}$ axis and depolarizes. 
Physically, this corresponds to an effectively resonant excitation of (individual) spins that arises when one of the driving harmonics  is tuned close to their energy: $|h_i -  m^* \omega_0 | < \epsilon/\tau_1 $ for some optimal integer $m^*$ [see Fig.~\ref{fig:local}(b)].
Such resonances occur with a small probability in the limit of $\epsilon \ll W \tau_1$, and  amount to a reduction of the total polarization by a constant  factor proportional to $\epsilon/({\rm min}(W,\omega_0)\tau_1)$. However, if  $\epsilon \sim W\tau_1$, a substantial fraction of spins rapidly depolarize due to resonant processes shown in Fig.~\ref{fig:local}(b).
Note that the phenomenological phase boundary extracted in Ref.~\cite{Choi16DTC}, based on the existence of  self-consistent closed spin trajectories, is consistent with the perturbative condition  $\epsilon \ll W\tau_1$.

We next focus on the second-order process illustrated in Fig.~\ref{fig:local}(c) and (d) in which  a pair of spins $I$ and $J$ simultaneously flip their polarizations while exchanging their energies with each other \emph{and} with the external drive.
Such  processes are resonant when
\begin{align}
\label{eqn:pair_res}
\Delta_{IJ} \approx |h_I + h_J - m^*\omega_0| < J^\textrm{eff}_{IJ},
\end{align}
 where $h_I$ and $h_J$ are effective on-site potential for spins $I$ and $J$, respectively, $m^*$ is the  optimal harmonic number that minimizes the energy difference, and $J^\textrm{eff}_{IJ}$ is the effective amplitude of the pair-flip process.
The amplitude $J^\textrm{eff}_{IJ}$
\begin{align}
\label{eqn:effective_coupling}
J^\textrm{eff}_{IJ} \sim \left(\frac{\epsilon}{\tau_1} \right)^2 \frac{J_{IJ}}{ r_{IJ}^3}  \left(\frac{1}{\delta_{I}^2}+\frac{1}{\delta_{J}^2}\right),
\end{align}
can be estimated from the interference of two paths in the second-order perturbation theory, as illustrated in Figs.~1(c) and 1(d). Here  $\delta_{I(J)} \equiv {\rm min}_{\ell^*}(h_{I(J)} - \ell^* \omega_0)$ is the energy difference between initial or final states and intermediate virtual states, up to extra energy provided by a driving harmonic $\ell^*$. 
We find that $J^\textrm{eff}_{IJ}$ is an effective  long-range interaction decaying as $\sim 1/r_{IJ}^3$ allowing the flipping of remote spin pairs.

The resonance condition \eqref{eqn:pair_res} is sensitive to $\omega_0$ and behaves qualitatively differently in two limiting cases: (i) $\omega_0 \gg W$ and (ii) $\omega_0 \ll W$  [see Fig.~\ref{fig:local}(c,d)].
In the former case, the optimal choice is  $m^* = \ell^* = 0$ since spins cannot absorb or emit such a large energy $\omega_0$. In the latter case, effective energy differences (both $\Delta_{IJ}$ and $\delta_{I},\delta_J$) are bounded by $\omega_0$ as the external drive can always compensate energy in units of $\omega_0$. These considerations yield the scaling $J^\textrm{eff}(r) \sim C J_0 / r^3$ with
\begin{align}
C \approx 
\left\{
\begin{array}{cc}
(\epsilon/\tau_1 W)^2  &\textrm{for } \omega_0 \gg W\\
\epsilon^2  &\textrm{for }  \omega_0 \ll  W
\end{array}
\right.,
\end{align}
and the effective range $W^\textrm{eff}$ of the energy differences $\Delta_{IJ}$ becomes $W^\textrm{eff} \sim W$ for $\omega_0 \gg W$ and $W^\textrm{eff} \sim \omega_0$ for $\omega_0 \ll W$.


We now estimate the probability that a given spin finds a resonant partner within a ball of radius $R$. This is obtained by integrating the probability of finding such a partner in a shell $R$ and $R+dR$
\be\label{eq:decay_prob}
dP=\left(J^\textrm{eff} (R)/W^\textrm{eff}\right) n_0 4\pi R^2 dR,  
\ee
from a short distance cutoff $a_0$ to $R$, which gives $P(R) \sim \log  (R/a_0)$. Here the first factor in 
Eq.~(\ref{eq:decay_prob}) is the probability of satisfying Eq.~\eqref{eqn:pair_res}, and the second factor is the average number of spins within a shell of size  $R$ with the density $n_0$. As this probability diverges, it implies that pairwise spin flips prevail, and the system thermalizes, with the DTC order slowly decaying over time.
We can extract the time scale associated with these pair-spin-flip processes using the typical distances $R_*$ of resonant spin pairs.
%
%
%
Solving $P(R_*) \sim 1$ gives $R_* \approx  a \exp{[W^\textrm{eff}/4\pi C J_0 n_0]}$.
Finally, the effective depolarization rate is estimated from the interaction strengths of typical pairs, i.e., $\tilde{\Gamma} \sim J^\textrm{eff}(R_*) $, leading to the decay rate per Floquet cycle $\Gamma \equiv \tilde{\Gamma} \tau_1$:
\begin{align}
\Gamma \sim 
\left\{
\begin{array}{cc}
\frac{J_0\epsilon^2}{a_0^3\tau_1 W^2} 
 \exp {\left[- \frac{3W^3 \tau_1^2}{4\pi J_0n_0 \epsilon^2}\right]}&\textrm{for } \omega_0 \gg W \\
\frac{J_0\epsilon^2 \tau_1}{a_0^3} 
\exp {\left[- \frac{3 }{ 2 J_0n_0 \epsilon^2 \tau_1}\right]} & \textrm{for } \omega_0 \ll W.
\end{array}
\right.
\label{eqn:Gamma}
\end{align}
%
This exponentially slow in $1/\epsilon^2$ decay of the DTC order is a central result of the present Letter and is a direct consequence of critically slow thermalization of dipolar systems in 3D~\cite{anderson,kucsko}. Interestingly, the depolarization is exponentially sensitive to the parameters $\tau_1$ and $\epsilon$  in two distinct ways: In regime (i) $\Gamma$ is a function of $\tau_1^2/\epsilon^2$ while in regime (ii) it only depends on $1/\epsilon^2 \tau_1$. These considerations allow us to identify  an effective phase boundary using the criteria $\tau_1^2/\epsilon^2 = A$ or $1/\epsilon^2 \tau_1 = B$ with some constants $A$ and $B$.   Remarkably,  this boundary illustrated in Fig.~2  captures the key features observed in the experiment \cite{Choi16DTC}: the linear growth of $\epsilon$ for short $\tau_1$ and slow diminishing of $\epsilon$ at longer $\tau_1$~\cite{Corrections}.


\begin{figure}[t]
\center
\includegraphics[width=0.9\columnwidth]{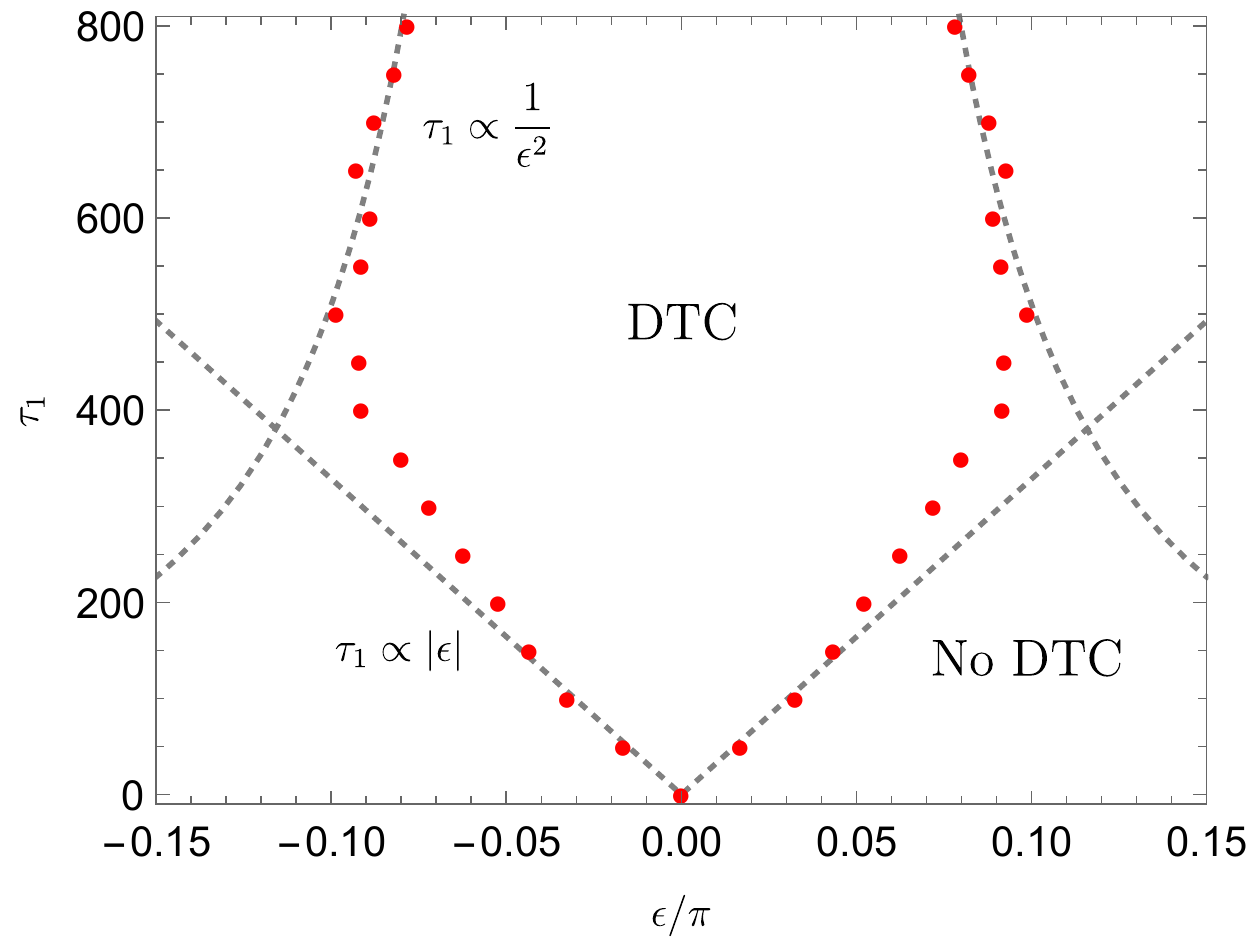}
\caption{Phase diagram of the DTC obtained numerically (see~\cite{SOM} for details). Dotted lines indicate limiting behaviors of the phase boundary: at high driving frequencies, the phase boundary is linear, $\tau_1 \propto |\epsilon|$, while for low driving  frequencies, it closes up as $\tau_1 \propto 1/\epsilon^2$, c.f.~Eq.~(\ref{eqn:Gamma}). This is in good agreement with the experimental observations of Ref.~\cite{Choi16DTC}.}
\label{fig:phase_diagram}
\end{figure}

\begin{figure}[t]
\center
\includegraphics[width=0.9\columnwidth]{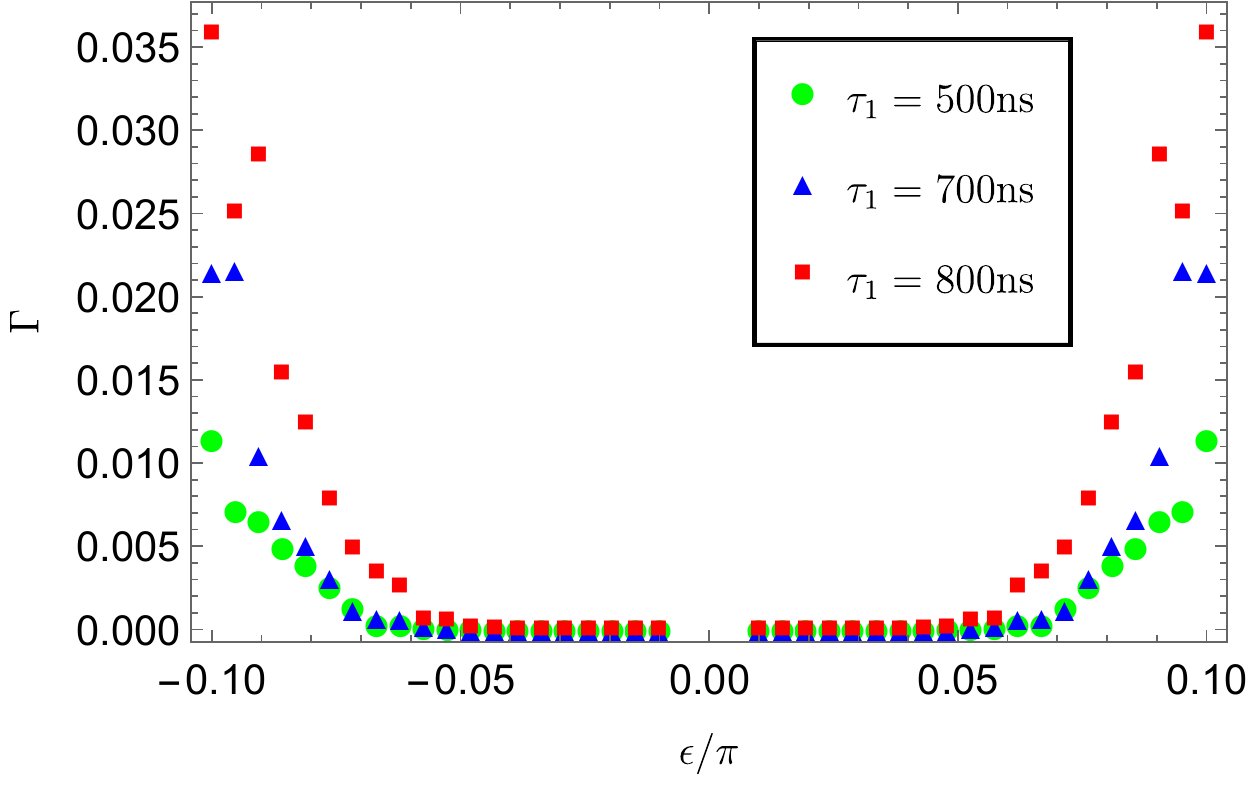}
\caption{Decay rate versus perturbation $\epsilon$ for various $\tau_1$s obtained numerically~\cite{SOM} . One sees a sharp rise of the decay rate as one crosses the DTC phase boundary (determined as the $\epsilon$ for which $\Gamma(\epsilon,\tau_1) = 1/100$), which is reminiscent of a phase transition.}
\label{fig:depolarization_rate}
\end{figure}

\emph{Technical procedure.}---We now outline the technical procedure that formalizes the above discussion (see~\cite{SOM} for details).
The key idea is to identify a time-dependent unitary transformation of the Hamiltonian $H(t)$ such that nonresonant single spin flips are essentially ``integrated out'' and only residual two-spin-flip processes become dominant terms in the effective Hamiltonian $H'(t)$.
More specifically, we start from the Hamiltonian~\eqref{eqn:H} with $H_0$ representing the Ising interactions and $V$ the applied field, and perform a time-periodic unitary transformation $Q(t+\tau_1)=Q(t)$, which gives rise to
\begin{align}
H'(t) & = Q(t)^\dagger  ( H_0 +  \epsilon V \sum_n \delta(t - n^{-}\tau_1) - i \partial_t  ) Q(t). 
\label{eqn:Hp}
\end{align}
Our goal is to eliminate terms that are linear in $\epsilon$ from $H'(t)$. Following Ref.~\cite{Abanin20161}, we look for $Q(t)$ of the form $Q(t) = e^{\epsilon \Omega(t)}$ with anti-Hermitian operator $\Omega (t)=\sum_n \Omega^{(n)}e^{in\omega_0t}$.
Expanding Eq.~($\ref{eqn:Hp}$) in powers of $\epsilon$, and requiring that the $O(\epsilon)$ term equals 0 gives an equation for  the $n$th Fourier mode $\Omega^{(n)}$: 
\begin{align}\label{eq:omega_n}
\frac{V}{\tau_1} - [\Omega^{(n)}, H_0] + n\omega_0 \Omega^{(n)}=0.
\end{align}
The matrix elements of the operator $\hat\Omega^{(n)}$ can be computed in the eigenstate basis $|s\rangle$ of $H_0$ (which is a product state basis in $S_i^x$ operators):
\begin{align}
 \langle s' | \Omega^{(n)} | s \rangle =  \frac{\langle s' | {V} | s \rangle}{(E_s - E_{s'} - n\omega_0)\tau_1 }. 
\label{eqn:Omega_matrix_elements}
\end{align}
Noting that ${V} = \sum_i S^y_i$, the operator $\Omega^{(n)}$ has nonzero matrix elements only between spin configurations $s$ and $s'$ that differ by one spin flip. If $| s \rangle $ and $| s' \rangle$ differ by the value of spin $I$, $E_s - E_{s'} = 2 \sum_{j \neq I}  \frac{J_{jI}}{r_{jI}^3} S_j^x(s) S_I^x(s)=2h_I S_I^x.$ We assume that the on-site field $h_I$ is random (due to positional disorder and orientation dependence of $J_{ij}$) and sufficiently strong such that resonances  are rare, i.e. the denominator in (\ref{eqn:Omega_matrix_elements}) typically does not diverge and the procedure controlled. 
Then, the rotated Hamiltonian to second order becomes
\begin{align}
H'(t) = H_0 - \frac{\epsilon^2}{2} [\Omega(t), V] \sum_n \delta(t-n^{-} \tau_1 )
\end{align}
A straightforward calculation~\cite{SOM, flip-flop} using expression (\ref{eqn:Omega_matrix_elements}) gives an effective Hamiltonian of the following form:
\begin{equation}
\label{eq:eff_hamiltonian}
H' (t)=H_0+\sum_{IJ} \frac{A_{IJ} J_{IJ} }{r_{IJ}^3} \left( S_{I}^+ S_J^+ + \text{H.c.} \right)  \sum_n \delta(t-n^-\tau_1),
\end{equation}
 where $S_I^+ \equiv (S_I^z + i S_I^y)/\sqrt{2}$ is the spin raising operator in $S_I^x$ basis for the spin $I$, and $A_{IJ}$ is the coefficient
\be\label{eq:A_I}
A_{IJ}\approx -2 S_I^x(s) S_J^x(s) \left(\frac{\epsilon}{\tau_1}\right)^2 \left(\frac{1}{\tilde\delta_I^2}+\frac{1}{\tilde\delta_J^2}\right), 
\ee
where we introduced the notation $
\frac{1}{\tilde\delta_J^2}=\sum_{\ell} \frac{1}{ \left( h_J - \ell \omega_0 \right)^2}.$ 

The effective Hamiltonian (\ref{eq:eff_hamiltonian}) contains the larger disordered part $H_0$, and long-range terms which can flip pairs of spins; the latter are suppressed proportional to $\epsilon^2$, leading to slow relaxation. From Eq.~(\ref{eq:A_I}) it is evident that the amplitudes for flipping a pair of spins depend on $h_I,h_J$, which in turn are determined by the positions of the spins. Assuming that $h_I,h_J$ take typical values of the order $W$, and taking the contribution of the harmonic $\ell^*$ for which  $h_J - \ell \omega_0$ is minimized (this gives the leading contribution to $\tilde\delta_J$), the expression (\ref{eq:A_I}) for the two-spin-flip amplitude reduces to the estimate (\ref{eqn:effective_coupling}) above.

We emphasize that the above unitary transformation is distinct from the rotating frame transformations employed to derive effective Hamiltonians in the high-frequency limit \cite{Abanin2017, AbaninPrethermal17}.
Rather, it utilizes the randomness of our Hamiltonian in order to effectively integrate out non-resonant single-spin-flip processes.

\emph{ Phase diagram.}---Using the effective Hamiltonian approach described above, we  obtain the phase diagram of the critical DTC. To improve upon the  estimates for $\Gamma (\epsilon, \tau_1)$, we take into account the fact that the distribution of the potential $h_i$ stems from the positional randomness of spins, and numerically sample $h_i$ from a distribution of $2000$ spins in a 3D region with density $9.26 \times 10^{-3} \text{~nm}^{-3}$ with a short distance cutoff of $3$~nm~\cite{kucsko}. 


While Eq.~(\ref{eqn:Gamma}) already provides analytical predictions for the decay rate $\Gamma$  by estimating the typical distance $R_*$ of resonant spin pairs, in numerics we find it more amenable to estimate $\Gamma$ from an explicit depolarization in time profile; the counting arguments in Eq.~\eqref{eq:decay_prob} predicts a power-law decay of polarization $q(n)$, from which the decay time scale $1/\Gamma$ is extracted by equating $q(n)$ to a small threshold~\cite{SOM}. The phase boundary is then identified from a criterion $\Gamma(\epsilon,\tau_1) = \Gamma_* =  1/100$. 

This approach yields the phase diagram illustrated in Fig.~\ref{fig:phase_diagram}, which is in very good agreement with  the experimental observations~\cite{Choi16DTC, PerturbativeRegion}. At high driving frequency, the boundary approximately follows a relation $\tau_1 \propto |\epsilon|$ (also obtainable using a semi-classical argument), while at low frequency $\tau_1 \propto 1/\epsilon^2$, which indicates that DTC order becomes less stable as $\tau_1$ is increased, due to the fact that multiphoton processes lead to faster depolarization. The DTC phase is most robust in the crossover regime, where $\omega_0 \sim W$. 

We also note that, strictly speaking, DTC order has finite relaxation rate at any $\epsilon\neq 0, \tau_1\neq 0$. However, we find that the relaxation rate $\Gamma$ increases very sharply at a certain value of $\epsilon$, as illustrated in Fig.~\ref{fig:depolarization_rate}, which matches the experimental observations and is reminiscent of a phase transition.  
Note, however, that unlike for a true phase transition, this increase does not become infinitely sharp even in the thermodynamic limit. 

\emph{Summary and discussion.}---We described a new approach to analyze the dynamics of periodically driven spin systems with long-ranged interactions and  applied it to explain the recently observed surprising stability of DTC in dipolar spin system. 
The results of our analysis are in very good agreement with experimental observations. They demonstrate that these observations correspond to a novel, critical regime of the DTC order. 

Furthermore, our general  approach can be applied to analyze the interplay of long-range interactions, randomness, and periodic driving in a broad class of experimental systems.  The present analysis focused on the experimentally relevant case of critical interactions, decaying as $1/r^\alpha$, where $\alpha$ coincides with the dimensionality of the system, $\alpha=d=3$. This leads to direct relaxation processes of spin pairs. It is interesting to extend the analysis to the case $\alpha>d$ (e.g. $\alpha=3, d=2$), where resonant spin-pair-flip processes are rare and presumably do not provide the main relaxation channel. Experimentally, such a situation can be realized by reducing the dimensionality of the dipolar spins systems. In the static case, relaxation is expected to occur via multispin processes: in essence, a sparse resonant network may form, which can act as a heat bath that mediates the relaxation of other spins~\cite{Burin06,Gutman16}. We expect that future experiments on DTC in reduced dimensions will allow one to probe such a delicate interplay of various relaxation mechanisms in driven systems with long-range interactions. Our theoretical approach is well suited for analyzing such systems. Finally, 
apart from these specific realizations, our analysis demonstrates  that the DTC response to periodic perturbations  can be used as a sensitive probe of nonequilibrium quantum states and phases of matter.

\begin{acknowledgements}
We thank I. Protopopov, V. Khemani, J. Choi, R. Landig, H. Zhou, A. Vishwanath and N. Yao for useful discussions. This work was supported   by Swiss National Science Foundation (W.W.H. and D.A.A), NSF, CUA, Vannever Bush Fellowship, ARO MURI, and Moore Foundation (S.C. and M.D.L.), and in part by the NSF under Grant No. NSF PHY11-25915 (W.W.H.,M.D.L., and D.A.A.). M.D.L., W.W.H. and D.A.A. are grateful to KITP, where this work was started, for hospitality during the program {\it Synthetic Quantum Matter}.
\end{acknowledgements}

\bibliography{mbl_time_crystal}

\begin{thebibliography}{27}
\bibitem [{Cor()}]{Corrections}%
  \BibitemOpen
  \href@noop {} {}\bibinfo {note} {To be precise, for short $\tau_1$ the phase
  boundary is not strictly linear but will contain a logarithmic correction:
  $\tau_1 \approx c_1 |\epsilon| \sqrt{\log{\left({c_2}/{\Gamma_*} \right) +
  \log|\epsilon|} } $, valid for $|\epsilon| \gg \Gamma_*/c_2$. The constants
  $c_1, c_2$ can be easily obtained by equating $\Gamma$ in
  Eq.~(\ref{eqn:Gamma}) to $\Gamma_* = 1/100$ and solving, and are functions of
  the physical parameters $J_0, n_0, W$.}\BibitemShut {Stop}%
\bibitem [{fli()}]{flip-flop}%
  \BibitemOpen
  \href@noop {} {}\bibinfo {note} {Note that we have neglected flip-flop terms
  $S^+_I S^-_J + h.c.$ since they are inconsequential for studying dynamics
  beginning from the initially polarized state in the $\hat x$ direction.
  Moreover, they do not give rise to depolarization as these terms are
  polarization conserving. See~\cite{SOM} for more discussion.}\BibitemShut
  {Stop}%
\bibitem [{Per()}]{PerturbativeRegion}%
  \BibitemOpen
  \href@noop {} {}\bibinfo {note} {Note that this phase boundary is consistent
  with the perturbative condition $\epsilon \ll W \tau_1$.}\BibitemShut {Stop}%
\end{thebibliography}

\section{Supplemental material: Critical time crystals in dipolar systems}

\section{A. Details on technical procedure}
In this section, we present the details on the technical procedure used to rotate the depolarization inducing Hamiltonian
\begin{align}
H(t) = \sum_{ij}  \frac{J_{ij} }{r_{ij}^3} S_i^x S_j^x + \epsilon \sum_i S^y_i \sum_n \delta(t-n^- \tau_1)
\label{eqn:originalH},
\end{align}
into an effective Hamiltonian $H'(t)$. 

As mentioned in the main text, because of the disorder in the interactions, single spin-flip processes, effected by the action of a single operator $S^y_i$ (i.e.~the off-diagonal $O(\epsilon)$ term in Eq.~(\ref{eqn:originalH})), are typically not resonant and do not induce significant depolarization. One has to consider other channels for depolarization which are of higher order in $\epsilon$, such as two, three, spin-flip processes, and inquire if they are resonant processes -- if so, then the dominant channel of decay is the one that governs the asymptotic behavior of depolarization dynamics. 

The purpose of the transformation we employ is thus to extract the terms that give rise to dominant depolarization processes: we will rotate the original Hamiltonian in such a way that off-diagonal terms that generate non-resonant processes are `integrated out', giving a resulting effective Hamiltonian $H'(t)$ whose leading order off-diagonal terms generate resonant proceses. In our system, it will turn out that the dominant decay channel is given by two spin-flip processes. 

\subsection{Rotating the Hamiltonian}

To that end, let us perform this transformation in detail. We write the unitary time evolution operator $U(t) = \mathcal{T} e^{-i\int_0^t H(t') dt'}$, generated by Eq.~(\ref{eqn:originalH}), as 
\begin{align}
U(t) = Q(t) \tilde{U}(t) Q^\dagger(0),
\end{align}
where we have yet to define the unitary $Q(t)$. With this decomposition, $\tilde{U}(t)$ is given by
\begin{align}
\tilde{U}(t) = \mathcal{T} e^{-i \int_0^t H'(t) dt},
\end{align}
i.e. it is generated by a rotated time-dependent Hamiltonian (via the Schrodinger equation):
\begin{align}
H'(t) & = Q(t)^\dagger \left(H_0 +  \epsilon V \sum_n \delta(t - n^{-} \tau_1) - i \partial_t \right) Q(t),
\label{eqn:Hp}
\end{align}
where
\begin{align}
& H_0 = \sum_{ij}  \frac{J_{ij}}{r_{ij}^3} S_i^x S_j^x, \\
& V = \sum_i S^y_i. 
\end{align}
Here $J_{ij} = J_0 q_{ij}$ where $q_{ij} = -(1-3 (\hat{z}\cdot \hat{r}_{ij})^2)$ encodes the angular dependence of the interactions between spins $(i,j)$, as in Ref.~\cite{Choi16DTC}.

We will pick $Q(t)$ to be time-periodic; then because $Q(n \tau_1) = Q(0)$, the expected value of observables as a function of time, such as the polarization (of one site) $S_i^x(n \tau_1)$, is given by
\begin{align}
& \langle \psi | S_i^x(n \tau_1) | \psi \rangle =\langle \psi | U^\dagger(n \tau_1) S_i^x U(n  \tau_1) | \psi \rangle \nonumber \\
& ~~~ = \langle \tilde{\psi} | \tilde{U}^\dagger(n \tau_1) \tilde{S}^x_i \tilde{U}( n\tau_1) | \tilde{\psi} \rangle, \nonumber \\
&  | \tilde{\psi} \rangle = Q^\dagger(0) |\psi \rangle \nonumber \\
& \tilde{S}^x_i = Q^\dagger(0) S^x_i Q(0),
\end{align}
where $| \psi \rangle$ is the initial state which we take to be polarized in the $x$-direction. In other words, $Q(0)$ is just some static rotation that rotates both the state and observable.

If in addition $Q(0)$ is a `small' rotation (as we will choose, and to be made precise below), then both the state and observable are close to the unrotated ones,$| \tilde{\psi} \rangle \approx | \psi \rangle$ and $\tilde{S}^x_i \approx S^x_i$, and one can conclude that
\begin{align}
\langle \psi | S_i^x(n \tau_1) | \psi \rangle \approx \langle \psi | \tilde{U}^\dagger(n \tau_1) S_i^x \tilde{U}(n \tau_1) | \psi \rangle.
\end{align}
That is, the time dependence (and consequently, depolarization) is completely captured in $\tilde{U}(t)$ and hence, $H'(t)$, the rotated effective Hamiltonian.

\subsection{Choosing the rotation $Q(t)$}
Now let us construct $Q(t)$, which we write as $Q(t) = e^{\epsilon \Omega(t)}$, where $\epsilon$ is explicitly the small parameter and $\Omega(t)$ a periodic anti-Hermitian operator. Expanding Eq.~($\ref{eqn:Hp}$) we get
\begin{align}
H'(t) &= H_0 + \epsilon \left(V \sum_n \delta(t-n^- \tau_1) - [\Omega, H_0] - i \partial_t \Omega  \right) \nonumber \\
&+ \epsilon^2 \left(\frac{1}{2} [\Omega, [\Omega, H_0]] - [\Omega, V] \sum_n \delta(t-n^- \tau_1) + \frac{i}{2} [ \Omega, \partial_t \Omega] \right)   \nonumber \\
& \cdots.
\label{eqn:oldrotatedH}
\end{align}
Utilizing a procedure similar to Ref.~\cite{Abanin20161}, we equate the order $\epsilon$ piece to $0$ with the constraint that $\Omega(t)$ is time-periodic. Note that this transformation is distinct from the transformations employed in Refs.~\cite{Abanin2017, AbaninPrethermal17} to generate effective Hamiltonians in high-frequency driven systems. There, the small parameter was the inverse of the driving frequency $\omega$, but here, in anticipation that we will take into account the disorder in the interactions, the small parameter is served by $\epsilon$, the strength of the off-diagonal perturbation to $H_0$. Decomposing $\Omega(t)$ in terms of its Fourier modes $\Omega (t)=\sum_n \Omega^{(n)}e^{in\omega_0t}$, where $\omega_0 = 2\pi/\tau_1$, and  using the fact the Fourier transform of the Dirac comb $\sum_n \delta(t-n^- \tau_1)$ is $\frac{1}{\tau_1} \sum_n e^{-i n \omega_0  t}$, the equation that the $n$-th Fourier mode has to obey is
\begin{align}
\frac{V}{\tau_1} - [\Omega^{(n)}, H_0] + \omega_0 n \Omega^{(n)} = 0 .
\end{align}
In the basis of the eigenstates $| s \rangle$  of $H_0$ which are product states in the $x$-direction, we can therefore write the solution as
\begin{align}
 \langle s' | \Omega^{(n)} | s \rangle =  \frac{\langle s' | {V} | s \rangle}{(E_s - E_{s'} - n\omega_0)\tau_1 }. 
\label{eqn:OmegaN}
\end{align}
Note that since $V = \sum_i S^y_i$, the eigenstates $s$ and $s'$ representing spin configurations in the matrix element of $\Omega^{(n)}$ can only differ by one spin-flip. If $| s \rangle $ and $| s' \rangle$ differ at the $I$th spin,
\begin{align}
E_s - E_{s'} = 2 \sum_{j \neq I} \frac{J_{jI}}{r_{jI}^3} S_j^x(s) S_I^x(s),
\end{align}
where $S_j^x(s)$ represents the $S^x$ component of the $j$th spin for the configuration $|s \rangle$, which for the starting state is just $+\frac{1}{2}$.

Because of the disorder in $J_{ij}/r_{ij}^3$ (accorded for by the random positions and relative angles between the spins), resonances (i.e. terms where the denominator $\approx 0$) are controlled, provided 
\begin{align}
\frac{\epsilon}{\tau_1} \ll \min_{m \in \mathbb{Z}} \left| \left(  2\sum_{i \neq J}  \frac{J_{iJ} }{r_{iJ}^3 } S_i^x(s) S_J^x(s)   - m \omega_0  \right) \right|.
\end{align}
This gives us a condition that our perturbative procedure should work only for
\begin{align}
\epsilon < \min(W,\omega_0) \tau_1,
\end{align}
where $W^2$ is the variance of the interactions,
\begin{align}
W^2 = \left\langle 4 \left( \sum_{i \neq J}  \frac{J_{iJ}}{r_{iJ}^3} S_i^x(s) S_J^x(s) \right)^2 \right\rangle.
\end{align}
The angular brackets represent averaging over different spins $i$, and we have used the fact that the mean is $0$.
In that case, the corresponding fraction of spins that are resonant is then small, and goes as  $\sim \frac{\epsilon}{\min(W,\omega) \tau_1}$.

\subsection{Effective Hamiltonian}
The rotated Hamiltonian (\ref{eqn:oldrotatedH}) then becomes
\begin{align}
H'(t) = H_0 - \frac{\epsilon^2}{2} [\Omega(t), V] \sum_n \delta(t-n^{-} \tau_1) + \cdots.
\label{eqn:rotatedH}
\end{align}
Let us concentrate on the second term and look at its matrix elements. This is 
\begin{align}
-\frac{\epsilon^2}{2 \tau_1} \langle s' | \sum_{n,m} [\Omega^{(m)}, V] | s \rangle e^{i \omega_0 (m+n) t}.
\end{align}
Now $\Omega^{(m)}$ and $V$ are sums of terms which each individually flip a single spin, so $|s \rangle$ and $|s' \rangle$ can differ by either only zero spin flips (i.e. $|s'\rangle = |s\rangle$) or two spin flips ($|s' \rangle = |s_{IJ} \rangle$) where $s_{IJ}$ stands for the spin configuration $s$ with the $I$-th and $J$-th spins flipped. The diagonal process (zero spin flips) leads to a renormalization of the energy of $H_0$, while the off-diagonal process where $V$ flips spin $I$ and $\Omega^{(m)}$ flips spin $J$ results in the matrix element
\begin{align}
& - \frac{\epsilon^2}{2 \tau_1} \sum_{m,n} \langle s_{IJ} | \Omega^{(m)}_J V_I - V_I \Omega^{(m)}_J | s \rangle e^{i \omega_0 (m+n) t} \nonumber \\
= & - \frac{\epsilon^2}{2 \tau_1} \sum_{m,k} e^{i \omega_0 k t} \left(  \langle s_{IJ}  | \Omega^{(m)}_J | s_I \rangle - \langle  s_{J} | \Omega_J^{(m)} | s \rangle  \right).
\end{align}
The term in the parenthesis is
\begin{align}
& - \frac{1}{(E_{s_I} - E_{s_{IJ}} -  m \omega_0 )\tau_1} + \frac{1}{(E_{s} - E_{s_{J}} -  m  \omega_0)\tau_1} \nonumber \\
= & - \frac{1}{(2 \sum_{i \neq I,J}   \frac{J_{iJ}}{r_{iJ}^3}  S_i^x(s) S_J^x(s) - 2   \frac{J_{IJ}}{r_{IJ}^3 } S_I^x(s) S_J^x(s)  -  m \omega_0  )\tau_1}   \nonumber \\
& + \frac{1}{(2 \sum_{i \neq J} \frac{J_{iJ} }{r_{iJ}^3 } S_i^x(s) S_J^x(s)   - m \omega_0 )\tau_1 } \nonumber \\
\approx & - \frac{4  \frac{J_{IJ}}{r_{IJ}^3} S_I^x(s) S_J^x(s) }{ \tau_1 \left( 2 \sum_{i \neq J}  \frac{J_{iJ} }{r_{iJ}^3 } S_i^x(s) S_J^x (s) - m \omega_0  \right)^2 },
\label{eqn:hopping}
\end{align}
where $\omega_0 = 2\pi/\tau_1$.

Using this, we can thus write the rotated Hamiltonian  (\ref{eqn:rotatedH}) in operator form:
\begin{align}
H'(t)& \approx H_0 + \sum_{IJ} \frac{A_{IJ} J_{IJ}}{r_{IJ}^3}(S_I^+ S_J^+ + h.c.) \sum_n \delta(t-n^- \tau_1) \nonumber \\
& + O(\epsilon^3),
\end{align}
where $S_I^+ \equiv S_I^y + i S_I^z$ is the spin-raising operator in the $S_I^x$ basis for spin $I$, $A_{IJ}$ is the coefficient of the interactions between two spins $(I,J)$, given by
\begin{align}
A_{IJ} &=  \left( \frac{\epsilon^2}{\tau_1^2}\right) \sum_m \frac{-2  S_I^x(s) S_J^x(s) }{ \left( 2 \sum_{i \neq J}  \frac{J_{iJ} }{r_{iJ}^3 } S_i^x(s) S_J^x (s) -  m  \omega_0 \right)^2 } \nonumber \\
& + (I \leftrightarrow J).
\end{align}
This effective Hamiltonian gives Eqs.~(5, 13) and (14) in the main text. Note that technically speaking, Eqs.~(\ref{eqn:hopping}) when written in operator form would also give flip-flop terms $S^+_I S^-_J + h.c.$. However, because we are ultimately interested in the dynamics of depolarization beginning from an initial polarized state  in the $\hat x$ direction,  and moreover such flip-flop terms are polarization conserving, they do not contribute to depolarization and so we drop them in the effective Hamiltonan.

Depending on the two limits of high or low frequencies ($\omega_0 \gg W$ or $\omega_0 \ll W$ respectively), $A_{IJ}$ has qualitatively different behaviors. Its scaling behavior with respect to $\epsilon$ and $\tau_1$ can be extracted in those limits, which is Eq.~(6) in the main text. Thus, this gives rise to different behaviors for the depolarization rate in the two limiting cases, and can ultimately be used to determine the phase boundary of the critical time crystal, as was done in the main text.

\section{B. Numerical procedure to extract decay rate and phase boundary}
In this section, we describe the numerical procedure used to 1) determine the decay rate $\Gamma$ as a function of the perturbation $\epsilon$, and 2) obtain the phase diagram of the critical DTC shown in the main text.

Assuming that the initial state is polarized in the $x$-direction, we first generated the probability distribution $P(h_I)$ of $ h_I$, which is the mean-field potential field felt by spin $I$, i.e.~$ h_I =   \frac{1}{2} \sum_{j \neq I} \frac{J_{Ij}}{r_{Ij}^3}$. Here $J_{ij} = J_0 q_{ij}$ and $q_{ij} = -(1- 3(\hat{z}\cdot \hat{r}_{ij})^2)$ encodes the angular dependence of two spins $(i,j)$. We modeled this by uniformly distributing $N = 2000$ spins in a 3-torus of linear dimension $L = 60$ nm, with a minimum (UV) distance cutoff $r_\text{UV} = 3$ nm, and caculating for each spin $h_i$. Such a choice of parameters gives a particle density of $n_0 = 9.26 \times 10^{-3}$ nm$^{-3}$. Using $J_0 = 2\pi \times 52$ Mhz nm$^{-3}$ as in the experiments, we can characterize this distribution by its variance $W^2$, which gives $W = 4.49$ Mhz.

From this distribution, we then drew two values $(h_I, h_J)$, provided that they are each not resonant. That is, we discarded values if
\begin{align}
\frac{\epsilon/\tau_1}{ | h_{I/J} - m \omega_0|} > \alpha_1  \text{ for all } m \in \mathbb{Z},
\end{align}
where $\alpha_1$ is some fixed $O(1)$ number. We determined this number from the condition that along the line $W \tau_1 = |\epsilon|$ where our perturbative analysis holds, at least $80\%$ of the spins are not resonant. In the region where $W \tau_1 > \epsilon$, this implies that $>80\%$ of the spins are off-resonant. In our simulation, $\alpha_1 = 3.7$. Then, $(h_I, h_J$) represent the mean-field potentials felt by spins $I$ and $J$ respectively.

Next, we estimated the probability that the two spins $(I,J)$ a distance $r$ apart form a resonant pair. In principle this is obtained by comparing the effective hopping $J_{\rm eff}(r) \equiv A_{IJ} J_{IJ}/r^3$, against the minimum quasi-energy gap $\Delta$. Here 
$
A_{IJ} J_{IJ} \sim -\left( \frac{\epsilon^2}{\tau_1^2}\right)  J_0 \overline{q_{IJ}} \sum_m \left(\frac{1}{( h_I - m \omega_0)^2 } + \frac{1}{(h_J - m \omega_0)^2 } \right)
$
and 
$
\Delta \approx \min_m \left| h_I + h_J-  m  \omega_0 \right|.
$
  The bar over the angular dependence $q_{ij}$ represents the typical angular dependence; this can be derived analytically and gives $\overline{q_{ij}} = 2/\sqrt{5}$.  In other words, for a fixed $r$, if
\begin{align}
 \frac{J_{\rm eff}(r)}{ \Delta} > \alpha_2, 
\label{eqn:prob}
\end{align}
where $\alpha_2$ is another order one number, then this spin pair is resonant. We take $\alpha_2 = 3.4$.

In practice, we make a simplifying assumption that for large enough $r$, which we call $d$, the probability of finding two spins a distance $r$ apart which are resonant is small and is simply proportional to ${J_{\rm eff}(r)}/{ \alpha_2 \Delta} \propto 1/r^3$ (we have numerically checked this assumption holds).

Then, one can simplify the counting of resonant pairs by just estimating the the probability $\mathcal{P}$ of finding two spins some fixed distance $d$ apart. We generate $4000$ pairs $(I,J)$, and count the number of pairs that satisfy Eqn.~(\ref{eqn:prob}) for $r= d$; the fraction of such pairs is $\mathcal{P}$. To get the probability of resonant pairs at distances $R > d$, we can then just multiply $\mathcal{P}$ by the factor $(d/r)^3$. In the simulations, we obtain $d$ as the distance for which the last step of Eqn.~\ref{eqn:hopping} is justified, that is, if $W \sim J_0 \overline{q_{ij}}/d^3$, which gives $d = 4.02$ nm for the parameters we have used.

Having determined $\mathcal{P}$, we extracted the decay rate $\Gamma(\epsilon,\tau_1)$. The survival of polarization probability is given by a power law $P(t) =  (t/t_0)^{-q}$  where $q \propto \mathcal{P}$. This can be derived as the product of probabilities of having no resonant spins at each distance $[r, r+ dr]$ up to $R(t) = (J_0 t)^{1/3}$  \cite{Choi16DTC}:
\begin{align}
P(t) &= \prod_{r=r_\text{UV}}^{R(t)} \left( 1 - 4\pi n_0 r^2 dr \frac{\overline{A_{IJ} J_{IJ} } }{d^3 \alpha_2 \Delta } \frac{d^3}{r^3} \right) 
= \left(t/t_0 \right)^{-q},
\end{align}
where
\begin{align}
q &= \frac{4\pi n_0 d^3 }{3 }  \left( \frac{\overline{A_{IJ} J_{IJ} } }{\alpha_2 \Delta d^3}\right), \nonumber \\
t_0 &= \frac{r_\text{UV}^3}{J_0}.
\end{align}
The factor in the parenthesis $\left( \frac{\overline{A_{IJ} J_{IJ} } }{\alpha_2 \Delta d^3}\right)$ is the probability $\mathcal{P}$, and hence we see $q \propto \mathcal{P}$.

 The dimensionless decay rate per Floquet cycle $\Gamma$ can then be obtained from $\Gamma(\epsilon,\tau_1) = 1/n^*(\epsilon,\tau_1)$ where $n^*(\epsilon,\tau_1)$ is the number of Floquet cycles such that $P(t)$ drops to some fixed threshold $1/A$. Solving $P(n^* \tau_1) = 1/A$ for $n^*$ yields
\begin{align}
\Gamma(\epsilon,\tau_1) = \frac{\tau_1}{t_0} e^{-\beta/\mathcal{P}},
\end{align}
where 
\begin{align}
\beta = \frac{3 \log(A)}{4\pi n_0 d^3}.
\end{align}
We take $A = 10$; physically this corresponds to the situation where the polarization drops to $10\%$ of its starting value.

Lastly, to get the phase diagram, we estimated the phase boundary of the time crystal as the contour in the $\epsilon$-$\tau_1$ plane satisfying $\Gamma(\epsilon,\tau_1) = \Gamma_* =  1/100$. In other words, the phase boundary demarcates the regions in $\epsilon$-$\tau_1$ parameter space having significant decay after $100$ Floquet cycles or not. The choice of $100$ Floquet cycles was picked to match the experimental observations. Referring to Fig.~2 in the main text, one sees the linear ($ \tau_1 \propto |\epsilon|$) phase boundary for small $\tau_1$, and also the `closing up' of the phase boundary, $\tau_1 \propto 1/\epsilon^2$, for large $\tau_1$, as predicted by our  theory.

\end{document}